\documentclass[twocolumn]{webofc}
\usepackage[varg]{txfonts}   

\newcommand{\rts}{\sqrt{s_{\rm _{NN}}}}
\newcommand{\tF}{\tau_{\rm _F}}
\newcommand{\AT}{A_{\rm T}}
\newcommand{\dETdy}{\frac{dE_{\rm T}}{dy}}
\newcommand{\dNBdy}{\frac{dN_{\rm netB}}{dy}}
\newcommand{\ep}{\epsilon}
\newcommand{\nB}{n_{\rm _B}}
\newcommand{\nQ}{n_{\rm _Q}}
\newcommand{\nS}{n_{\rm _S}}
\newcommand{\muB}{\mu_{\rm _B}}
\newcommand{\muQ}{\mu_{\rm _Q}}
\newcommand{\muS}{\mu_{\rm _S}}

\begin{document}

\title{Calculating QCD Phase Diagram Trajectories of Nuclear Collisions using a Semi-analytical Model}
\author{
	\firstname{Todd} \lastname{Mendenhall} \inst{1}
	\and
	\firstname{Zi-Wei} \lastname{Lin} \inst{1}
}
\institute{
Department of Physics, East Carolina University, Greenville, NC 27858}

\abstract{
At low to moderate collision energies where the parton formation time $\tF$
is not small compared to the nuclear crossing time, the finite nuclear thickness
significantly affects the energy density $\ep(t)$ and net
conserved-charge densities such as the net-baryon density $\nB(t)$
produced in heavy ion collisions. As a result, at low to moderate 
energies the trajectory in the QCD phase diagram is also affected by
the finite nuclear thickness. Here, we first discuss our semi-analytical model
and its results on $\ep(t)$, $\nB(t)$, $\nQ(t)$, and $\nS(t)$ in
central Au+Au collisions. We then compare the $T(t)$, $\muB(t)$,
$\muQ(t)$, and $\muS(t)$ extracted with the ideal gas equation of
state (EoS) with quantum statistics to those extracted with a lattice
QCD-based EoS. We also compare the $T-\muB$ trajectories with the
RHIC chemical freezeout data. Finally, we discuss the effect of
transverse flow on the trajectories.}

\maketitle

\section{Introduction}
\label{sec-1}

The Bjorken energy density formula~\cite{Bjorken:1982qr} predicts the energy
density in the central spacetime rapidity region produced in the initial state of
heavy ion collisions assuming that partons originate at
$(z_0,t_0)=(0,0)$. After averaging over the transverse overlap area
$\AT$ (where we take the radius $R_A=1.12A^{1/3}$ fm from the
hard-sphere model), one needs to take a finite initial time because
the Bjorken energy density formula diverges as $t\to 0$:
\begin{equation}
\epsilon^{_{Bj}}(t)=\frac{1}{t\,\AT}\dETdy.
\label{ep-bj}
\end{equation}
Here, $dE_{\rm T}/dy$ is the transverse energy rapidity density at mid-rapidity.
A similar formula can be used to calculate the net-baryon density as a
function of time~\cite{Mendenhall:2021maf}:
\begin{equation}
\nB^{_{Bj}}(t)=\frac{1}{t\,\AT}\dNBdy, 
\label{nB-bj}
\end{equation}
which depends on the net-baryon rapidity 
density at mid-rapidity $dN_{\rm netB}/dy$. 
In Eqs.~\eqref{ep-bj}-\eqref{nB-bj}, the peak density 
occurs at the earliest time, which we take as 
the parton proper formation time $\tF$.

In our semi-analytical model~\cite{Mendenhall:2020fil}, we neglect
secondary parton interactions and consider that produced partons are
free-streaming, like the Bjorken energy density formula of Eq.~\eqref{ep-bj}.
However, we include the finite nuclear thickness by considering the
finite time $x$ and longitudinal width $z_0$
of the primary NN collisions, and obtain for the energy density:
\begin{equation}
\epsilon(t)=\frac{1}{\AT}
\iint_S\frac{dx\,dz_0}{t-x}\frac{d^3m_{\rm _T}}{dx\,dz_0\,dy}\cosh^3{\!y}.
\label{ep-dens}
\end{equation}
We then simplify the above integral 
by assuming that $d^3m_{\rm _T}/(dx\,dz_0\,dy) \propto dm_{\rm
  _T}/dy$, i.e., the initial transverse mass rapidity density is
uniformly distributed over the initial production area
$S$ in the $x-z_0$ plane~\cite{Mendenhall:2020fil}. 
Note that $dm_{\rm _T}/dy=dE_{\rm T}/dy+m_{\rm _N}dN_{\rm netB}/dy$
where $m_{\rm _N}$ is the nucleon mass.
Recently, we have further extended our semi-analytical
model~\cite{Mendenhall:2021maf} to calculate the net conserved-charge
densities including the net-baryon density $\nB(t)$ as
\begin{equation}
\nB(t)=\frac{1}{\AT}\iint_S\frac{dx\,dz_0}{t-x}\frac{d^3N_{\rm netB}}{dx\,dz_0\,dy}\cosh^2{\!y}.
\label{nB-dens}
\end{equation}
Since the initial net-charge comes from incoming protons and 
there is no net-strangeness in the incoming nuclei, 
the net-electric charge and net-strangeness densities in our semi-analytical
model are respectively given by 
\begin{equation}
\nQ(t)=\nB(t)\frac{Z}{A}\text{, and }\nS(t)=0.
\label{nQ-nS}
\end{equation}

Using the densities from our semi-analytical model, 
the temperature $T(t)$ and chemical potentials $\mu(t)$ can then be
extracted for the ideal gas EoS with quantum statistics 
with the following relations~\cite{Mendenhall:2021maf}:
\begin{align}
&\epsilon=\frac{19\pi^2}{12}T^4+3\frac{(\muB-2\muS)^2+\muS^2}{2}T^2
+3\frac{(\muB-2\muS)^4+\muS^4}{4\pi^2}, \nonumber \\
&\nB=\frac{\muB-\muS}{3}T^2+\frac{(\muB-2\muS)^3+\muS^3}{3\pi^2}, \nonumber \\
&\nQ=\frac{2\muB-5\muS}{3}T^2+\frac{2(\muB-2\muS)^3-\muS^3}{3\pi^2}.
\end{align}
In the above, we assume that the quark-gluon plasma (QGP) consists of
massless gluons and quarks, and we have used the fact that $\nS(t)=0$
for the ideal gas EoS leads to $\muQ=\muB-3\muS$. Therefore, the
problem of extracting a  $T-\muB$ trajectory in the QCD phase diagram
is reduced from solving a system of four equations with four unknowns
to solving the above system of three equations with three unknowns. 

One can also use a lattice QCD-based EoS to extract the
$T,\muB,\muQ,\muS$ from $\epsilon,\nB,\nQ,\nS$, where these quantities
are related with the standard thermodynamic
relations~\cite{Noronha-Hostler:2019ayj}. Each of the conserved-charge 
densities $n$ and the entropy density $s$ are given by a derivative of
the pressure $p$:
\begin{equation}
\begin{split}
\frac{\epsilon}{T^4}&=
\frac{s}{T^3}-\frac{p}{T^4}+\frac{\muB}{T}\frac{\nB}{T^3}
+\frac{\muQ}{T}\frac{\nQ}{T^3}+\frac{\muS}{T}\frac{\nS}{T^3}, \\
\frac{\nB}{T^3}&=
\frac{1}{T^3}\frac{\partial p}{\partial \muB}\Bigg\rvert_{T,\muQ,\muS},~
\frac{\nQ}{T^3}=
\frac{1}{T^3}\frac{\partial p}{\partial \muQ}\Bigg\rvert_{T,\muB,\muS}, \\
\frac{\nS}{T^3}&=
\frac{1}{T^3}\frac{\partial p}{\partial \muS}\Bigg\rvert_{T,\muB,\muQ},~
\frac{s}{T^3}=
\frac{1}{T^3}\frac{\partial p}{\partial T}\Bigg\rvert_{\muB,\muQ,\muS}.
\end{split}
\label{lattice_EOS}
\end{equation}
In the above, the pressure $p$ is defined by a Taylor series in powers
of $\mu/T$ up to total power $i+j+k\leq 4$:
\begin{equation}
\frac{p(T,\muB,\muQ,\muS)}{T^4}=\sum_{i,\,j,\,k}\frac{1}{i!\,j!\,k!}\chi_{ijk}^{BQS}
\left(\!\frac{\muB}{T}\!\right)^{\! i}
\left(\!\frac{\muQ}{T}\!\right)^{\! j}\left(\!\frac{\muS}{T}\!\right)^{\! k},
\end{equation}
where the coefficients $\chi_{ijk}^{BQS}$ are parameterized as
functions of $T$~\cite{Noronha-Hostler:2019ayj} based on lattice QCD
results.

\section{Results}
\label{sec-2}

\begin{figure}
\centering
\includegraphics{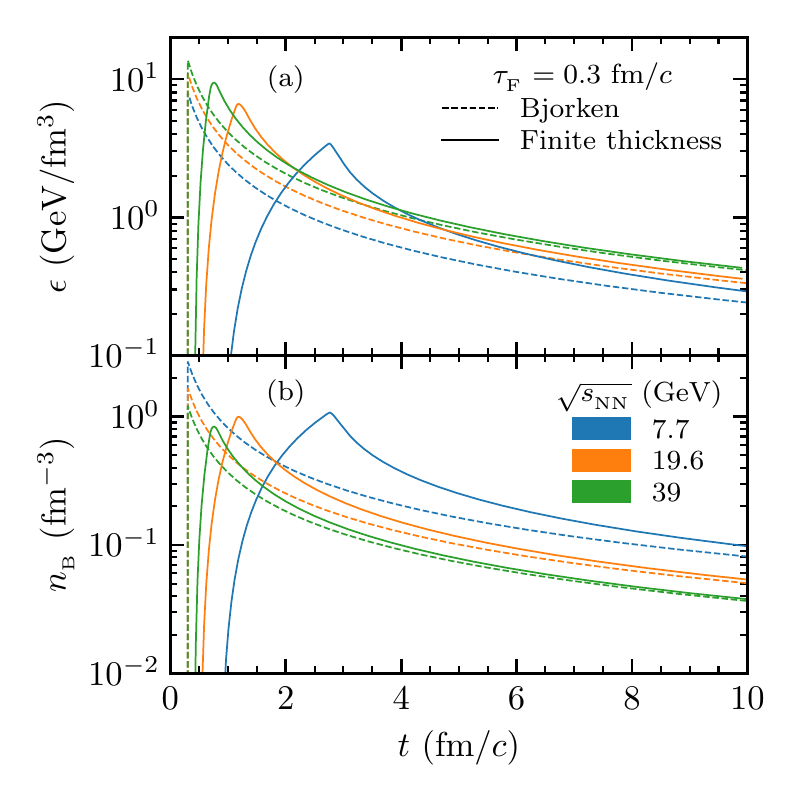}
\caption{(a) $\epsilon(t)$ and (b) $\nB(t)$ at mid-spacetime-rapidity
  from the Bjorken formula (dashed) and our formula (solid) for
  central Au+Au collisions at $\rts=$ 7.7, 19.6, and 39 GeV with
  $\tF=$ 0.3 fm/$c$.}
\label{fig-1}
\end{figure}

\begin{figure}
\centering
\includegraphics{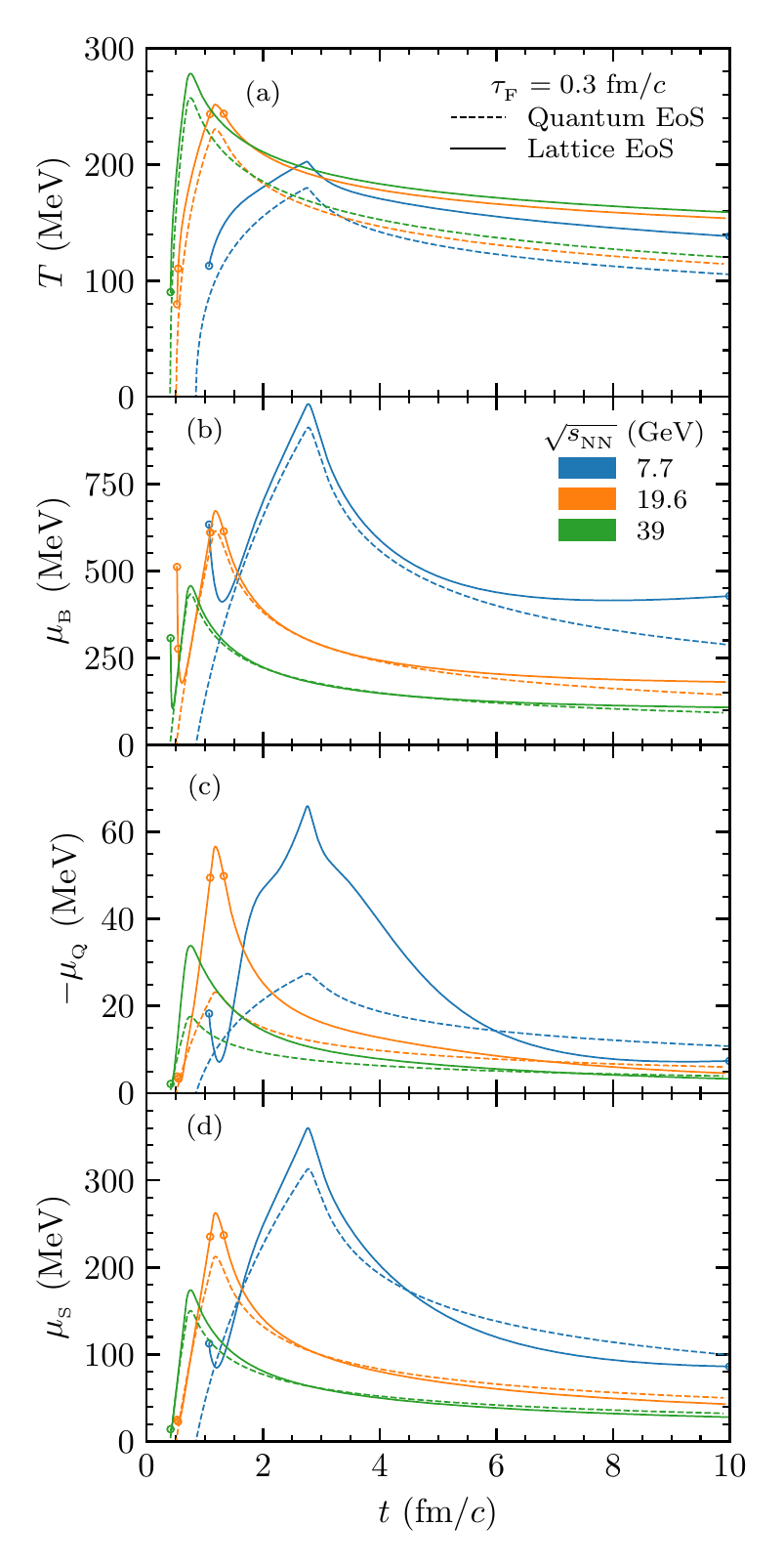}
\caption{(a) $T(t)$, (b) $\muB(t)$, (c) $-\muQ(t)$, and (d) $\muS(t)$
  extracted with the quantum EoS (dashed) and the lattice EoS (solid)
  for central Au+Au collisions at $\rts=$ 7.7, 19.6, and 39 GeV with
  $\tF=$ 0.3 fm/$c$. Open circles on the lattice EoS curves 
represent times when $\muB/T>2.5$.}
\label{fig-2}
\end{figure}

\begin{figure*}
\centering
\includegraphics{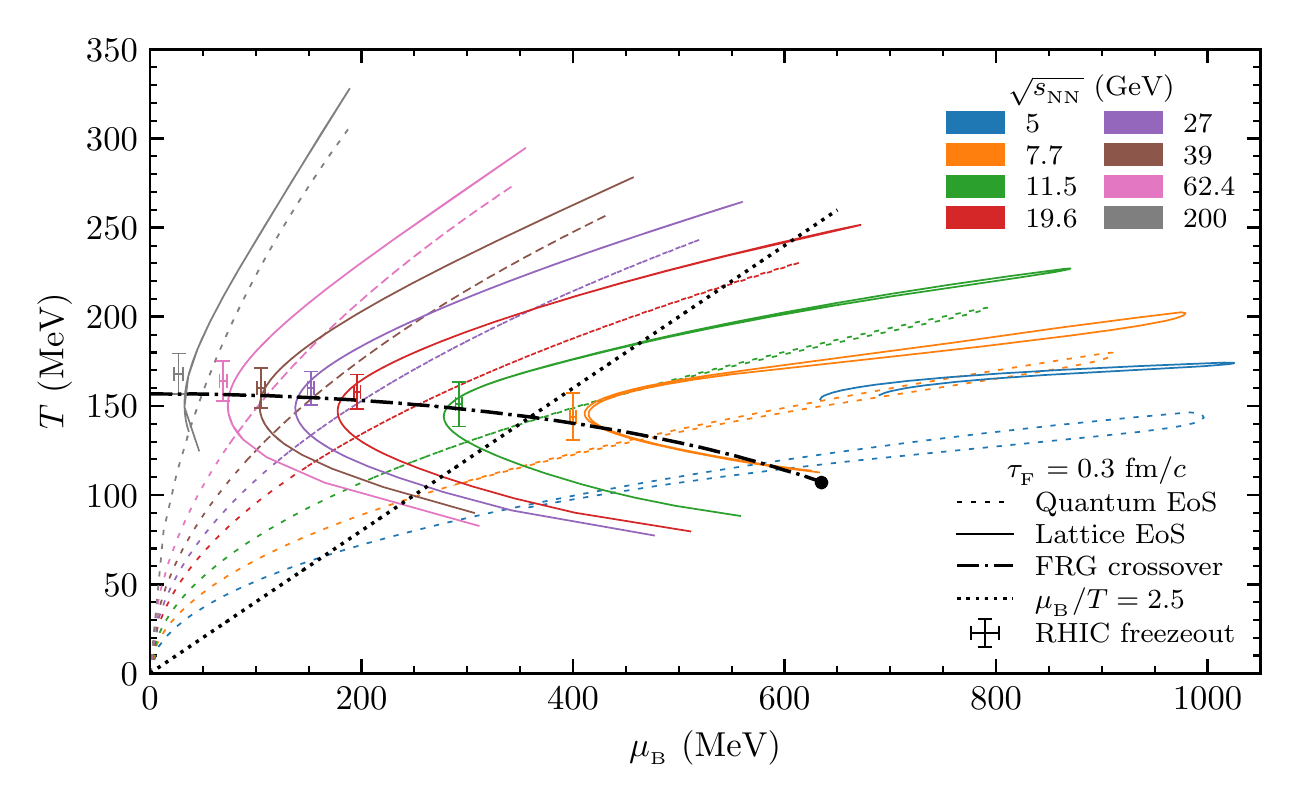}
\caption{QCD phase diagram trajectories extracted with the quantum EoS
  (dashed) and the lattice EoS (solid) compared with the RHIC chemical
  freezeout data (symbol with error bars) for central Au+Au collisions
  at $\rts=$ 5.0, 7.7, 11.5, 19.6, 27, 39, 62.4, and 200 GeV with
  $\tF=$ 0.3 fm/$c$. The FRG crossover curve (dot-dashed) with the 
  CEP and the $\muB/T=2.5$ line are also shown for reference.}
\label{fig-3}
\end{figure*}

In Fig.~\ref{fig-1}, we show the time evolution of the energy density $\ep(t)$
and net-baryon density $\nB(t)$ calculated with
Eqs.~\eqref{ep-dens}-\eqref{nB-dens} from our semi-analytical model
compared to those calculated with the Bjorken formulas of
Eqs.~\eqref{ep-bj}-\eqref{nB-bj}. The results are for central Au+Au
collisions at $\rts=$ 7.7, 19.6, and 39 GeV for a parton formation
time of $\tF=$ 0.3 fm/$c$. As $\rts$ increases, the maximum energy
density $\ep^{\rm max}$ in Fig.~\ref{fig-1} increases while the
maximum net-baryon density $\nB^{\rm max}$ decreases according to our 
semi-analytical model. While $\ep^{\rm max}$ also increases with
$\rts$ according to the Bjorken formula, it occurs at $t=\tF$ whereas
$\ep^{\rm max}$ from our semi-analytical model occurs later, at time
$t\in[t_a,t_2+\tF]$~\cite{Mendenhall:2021maf}. 
Note that $t_1$ and $t_2$ represent the starting and ending time of
the nuclear overlap, respectively; the nuclear crossing time is
$d_t=2R_{\rm   _A}/(\beta \gamma)$, and we choose $t_1=d_t/6$ and
$t_2=5d_t/6$~\cite{Mendenhall:2021maf}.

We also see in Fig.~\ref{fig-1} that for both the Bjorken formula and
our semi-analytical model, $\nB^{\rm max}$ decreases with
$\rts$ and it is reached at the same time as $\ep^{\rm max}$.
One major difference between the Bjorken formulas and our
semi-analytical model is that our densities start at zero (at 
$t=t_1+\tF$), increase to their maximum values, then decrease
thereafter. On the other hand, Bjorken densities start at their
maximum values and decrease with time. 
We also find that the late time evolution of our densities approaches
that of the Bjorken formula. This occurs because the formed partons in
our model must have $y\sim0$ in order to contribute to the densities in the
mid-spacetime-rapidity region at late times, just like the Bjorken
formula. 

Figure~\ref{fig-2} shows the time evolution of the temperature $T(t)$
and chemical potentials $\mu(t)$ for the quantum
EoS and the lattice EoS extracted using our densities at $\rts=$ 7.7,
19.6, and 39 GeV with $\tF=$ 0.3 fm/$c$. For both equations of state,
we extract the $T,\muB,\muQ$, and $\muS$ using the conditions in
Eq.\eqref{nQ-nS} from our semi-analytical model, which are relevant
for heavy  ion collisions and have also been used to constrain the
lattice  EoS~\cite{Noronha-Hostler:2019ayj}. As $\rts$ increases, the
maximum temperature $T^{\rm max}$ increases, but the maximum baryon
chemical potential $\muB^{\rm max}$ decreases in Fig.~\ref{fig-2}.
The results for the lattice EoS show that $\muB$ first decreases
before increasing with time, because the lattice EoS smoothly merges
with the hadron resonance gas model at 
$T\lesssim 135$ MeV~\cite{Noronha-Hostler:2019ayj}. The open circles
in Fig.~\ref{fig-2} represent the times when the lattice trajectories
are inside the region $\muB/T > 2.5$, where the lattice EoS is
expected to break down~\cite{Noronha-Hostler:2019ayj}.
We also observe in Fig.~\ref{fig-2}(c) that $\muQ^{\rm max}$ from the lattice
EoS can be much larger than that from the quantum EoS (by a factor
$\sim2$), while in Fig.\ref{fig-2}(d) the $\muS^{\rm max}$  values
extracted from the two EoS are reasonably close (within $\sim20\%$ of
each other). Note that a recent work~\cite{Wang:2021owa}
using the AMPT model, which includes secondary parton interactions, 
found similar results for the time dependences of $T$ and $\mu$ as our
results here.

In Fig.~\ref{fig-3}, we show the trajectories extracted from our
densities using the quantum and lattice equations of state 
in comparison with the RHIC chemical freezeout data, which were
obtained from grand canonical fits to the particle
yields~\cite{STAR:2017sal}. Trajectories for energies  
$\rts=$ 7.7, 11.5, 19.6, 27, 39, 62.4, and 200 GeV with $\tF=$ 0.3
fm/$c$ cross the crossover curve and can thus be compared with the
freezeout data, while the $\rts=5.0$ GeV lattice trajectory indicates a
problem in finding the full $T-\muB$ solution.  
We can see the effect of using the more realistic
lattice EoS on the extracted trajectories; e.g., the intersections
with the crossover curve from the functional renormalization group
(FRG)~\cite{Fu:2019hdw} shift to smaller $\muB$ 
and are closer to the RHIC chemical freezeout data.

The
maximum temperature reached by the trajectories 
extracted with the lattice EoS are also larger than that extracted
with the quantum EoS, which is also shown in Fig.~\ref{fig-2}. 
In addition, $\muB^{\rm max}$ extracted with the lattice
EoS is significantly larger ($\sim 10\%$) at low to moderate collision
energies than that with the quantum EoS. As $\rts$  
increases, the difference in $\muB^{\rm max}$ between the two equations of state
becomes smaller such that there is essentially no difference at
$\rts=200$ GeV. Note that the lattice trajectories at late times below
the FRG crossover curve do not approach the origin but instead go to a
finite $\muB$ and low $T$ in the QCD phase diagram.  
This behavior can also be seen in Fig.~\ref{fig-2}(b), 
where $\muB$ at late times can increase when using
the lattice EoS but always decreases when using the quantum EoS. 

In order to extract the lattice EoS trajectories in Fig.~\ref{fig-3}, we
calculate the intersection points between the constant $\epsilon$ and
$\nB$ contours in the $T-\muB$ plane that correspond to the $\epsilon(t)$
and $\nB(t)$ values at a given time $t$ from our
model~\cite{Mendenhall:2021maf}. We find that the lattice EoS does
not have $T-\muB$ solutions for low collision energies at very early
or very late times; this usually happens when the trajectory is in the
large $\muB/T$ region where the lattice EoS is expected to be
unreliable~\cite{Noronha-Hostler:2019ayj}.  
For example, the $\rts=5.0$ GeV lattice trajectory in
Fig.~\ref{fig-3} has no $T-\muB$ solution below the FRG crossover
line for the densities from our semi-analytical model. Moreover, we
find that no solution exists in the lattice EoS for our densities
at any time in the evolution of central Au+Au collisions at $\rts=2.0$
GeV~\cite{Mendenhall:2021maf}.  Therefore, the lattice EoS is an 
improvement over the ideal gas EoS at high collision energies where
$\muB/T<2.5$; however, it is expected to be unreliable at low
energies.

\begin{figure}
\centering
\includegraphics{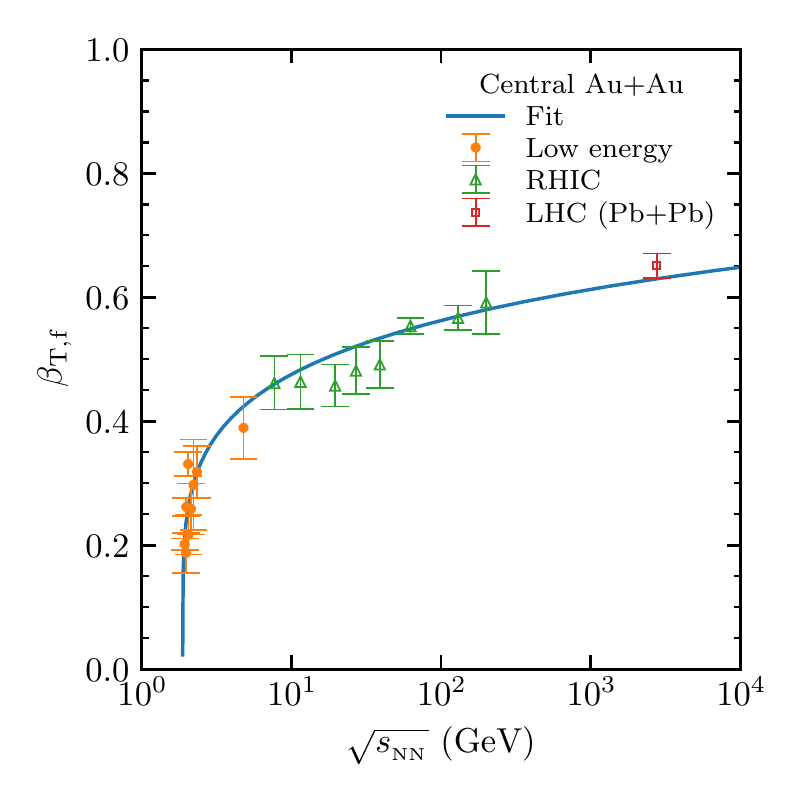}
\caption{Our parameterization (solid) of the final transverse velocity
  $\beta_{\rm _{T,f}}$ in central Au+Au collisions as a function of
  collision energy compared to kinetic freezeout values 
from multiple low energy experiments (circles), the Beam Energy
Scan program at RHIC (triangles), and the LHC (square).}
\label{fig-4}
\end{figure}

We have also investigated the effect of transverse expansion by
increasing the transverse overlap area $\AT$ with time $t$ in
Eqs.~\eqref{ep-dens}-\eqref{nB-dens}. The transverse radius $R_{\rm
  _T}$ of the overlap region increases according to a data-based
parameterization of the transverse flow velocity $\beta_{\rm _T}(t)$. 
We assume that $\beta_{\rm _T}(t)$ starts at 0 and
smoothly approaches a final value $\beta_{\rm _{T,f}}$. Using the
kinetic freezeout data, which were obtained by fitting the
transverse momentum spectra of central Au+Au collisions at various
collision energies~\cite{STAR:2017sal} to a blast-wave model, we
parameterize $\beta_{\rm _{T,f}}$ as~\cite{Mendenhall:2021maf}:
\begin{equation}
\beta_{\rm _{T,f}}=\left[\frac{\ln\left(\rts/E_0\right)}{64.7 +
    \ln\left(\rts/E_0\right)}\right]^{0.202}, 
\end{equation}
where $E_0=2m_{\rm _N}$ is the threshold energy. Figure~\ref{fig-4}
shows the parameterization in comparison with the kinetic freezeout
data~\cite{STAR:2017sal}. 
The data at low energies (orange circles) and RHIC energies (green
triangles) are for Au+Au collisions, while the data at the LHC energy
(red square) is for Pb+Pb collisions at 2.76 TeV (where we have
neglected the difference between Pb and Au for the kinetic freezeout
data).

Further details regarding the implementation and effects of
transverse expansion in our semi-analytical model can be found
in the full study~\cite{Mendenhall:2021maf}. Overall, we find that
including the transverse flow essentially does not change the path of
the trajectory (at a given $\rts$ and $\tF$), but it moves the
trajectory endpoint, the $T-\muB$ point corresponding to
$\epsilon^{\rm max}$ and $\nB^{\rm   max}$, a bit closer to the
origin. Importantly, the transverse expansion significantly decreases
the time spent in the parton phase (i.e., the QGP lifetime) at all
collision energies, and we also find that the QGP lifetime may 
have a local maximum below $\rts\sim11.5$
GeV~\cite{Mendenhall:2021maf}.

\section{Summary and Outlook}
\label{sec-3}

In this proceeding, we have calculated the $T-\muB$ trajectories in
the QCD phase diagram for central Au+Au collisions  using our
semi-analytical model, which includes the effect of the finite nuclear
thickness.  We have shown how the trajectories depend on the chosen
equation of state and that the trajectories extracted with a lattice
QCD-based EoS agree rather well with the chemical freezeout data from
the  RHIC Beam Energy Scan program. We also briefly discuss the 
implementation of transverse expansion and its effects on the
trajectories. We have written a web interface~\cite{Interface}, which
currently calculates the densities and trajectories after the user
specifies the colliding nuclei, $\rts$, $\tF$, and the ideal gas EoS
with quantum or Boltzmann statistics. We plan to further 
improve this web interface to include options to use the lattice EoS
and/or consider transverse expansion. We hope that our semi-analytical
model provides a useful tool for exploring the evolution of the dense
matter in the QCD phase diagram. \\

\begin{acknowledgement}
This work has been supported by the National Science Foundation under
Grant No. PHY-2012947.
\end{acknowledgement}

\bibliography{sqm_proceeding}

\begin{thebibliography}{8}

\bibitem{Bjorken:1982qr}
J.D. Bjorken, Phys. Rev. D \textbf{27}, 140 (1983)

\bibitem{Mendenhall:2021maf}
T.~Mendenhall, Z.W. Lin (2021), \texttt{2111.13932}

\bibitem{Mendenhall:2020fil}
T.~Mendenhall, Z.W. Lin, Phys. Rev. C \textbf{103}, 024907 (2021)

\bibitem{Noronha-Hostler:2019ayj}
J.~Noronha-Hostler, P.~Parotto, C.~Ratti, J.M. Stafford, Phys. Rev. C
  \textbf{100}, 064910 (2019)

\bibitem{Wang:2021owa}
H.S. Wang, G.L. Ma, Z.W. Lin, W.j. Fu, Phys. Rev. C \textbf{105}, 034912 (2022)

\bibitem{STAR:2017sal}
L.~Adamczyk et~al. (STAR), Phys. Rev. C \textbf{96}, 044904 (2017)

\bibitem{Fu:2019hdw}
W.j. Fu, J.M. Pawlowski, F.~Rennecke, Phys. Rev. D \textbf{101}, 054032 (2020)

\bibitem{Interface}
A web interface that performs our semi-analytical calculations is availabe at
  http://myweb.ecu.edu/linz/densities/  (2021)

\end{thebibliography}

\end{document}